\documentclass[12pt]{article}
\makeatletter
\def\fmslash{\@ifnextchar[{\fmsl@sh}{\fmsl@sh[0mu]}}
\def\fmsl@sh[#1]#2{%
  \mathchoice
    {\@fmsl@sh\displaystyle{#1}{#2}}%
    {\@fmsl@sh\textstyle{#1}{#2}}%
    {\@fmsl@sh\scriptstyle{#1}{#2}}%
    {\@fmsl@sh\scriptscriptstyle{#1}{#2}}}
\def\@fmsl@sh#1#2#3{\m@th\ooalign{$\hfil#1\mkern#2/\hfil$\crcr$#1#3$}}
\makeatother
\global\arraycolsep=2pt %reduces the separation in eqnarrays
\usepackage{bbm}
\usepackage{a4wide}
\usepackage{latexsym}
\usepackage{epsfig}

\def\beq{\begin{equation}}   \def\eeq{\end{equation}}
\def\bea{\begin{eqnarray}}   \def\eea{\end{eqnarray}}

\newcommand{\GeV}{\,\mbox{GeV}}

\begin{document} 
\thispagestyle{empty}
\rightline{TTP02-37}
\rightline{UND-HEP-02-BIG11}
\rightline{hep-ph/0212021}
\rightline{\today}
\bigskip
\boldmath
\begin{center}
{\bf \Large 
Parton--Hadron Duality in $B$ Meson Decays}
\end{center}
\unboldmath
\smallskip
\begin{center}
{\large{\sc Ikaros I. Bigi}\footnote{On leave from
         Physics Dept., University of Notre Dame du Lac,
   Notre Dame, IN 46556, U.S.A.}
   and {\sc Thomas Mannel}}
\vspace*{2cm} \\ 
{\sl Institut f\"{u}r Theoretische Teilchenphysik, \\
Universit\"{a}t Karlsruhe,  D--76128 Karlsruhe, Germany}\\[3cm]
{\it Contribution to the CKM workshop, \\
held at CERN, Geneva, Feb.15.-18.2002}
\end{center}
\vfill
\begin{abstract}
\noindent
We summarize the current view on Parton-Hadron duality
   as it applies to $B$ meson decays. It is emphasized that an OPE
   treatment is essential for properly formulating duality and its
   limitations. Duality violations are unlikely to become the limiting
   factor in describing semileptonic $B$ width vis-a-vie higher order
   corrections. The consistent extraction of the $b$ quark mass from
   $B$ production and decays provides a striking example of the
   theoretical control achieved.
\end{abstract}
\vfill
\newpage
%%%%%%%%%%%%%
\section{Introduction}
\label{INTRO}
%%%%%%%%%%%%%
%
%
%
Parton-hadron duality 
\footnote{This name might be more appropriate than the more frequently 
used {\em quark}-hadron duality since gluonic effects have to be 
included as well into the theoretical expressions. } 
-- or duality for short -- is one of the central 
concepts in contemporary particle physics. 
It is invoked to connect quantities evaluated on the 
quark-gluon level to the (observable) world of hadrons. 
It is used all the
time,  more often than not without explicit reference to
it. A striking example of the confidence the HEP community has in 
the asymptotic validity of duality was provided by the discussion 
of the width $\Gamma (Z^0 \to H_b H_b^{\prime}X)$. 
There was about a 2\% difference in the predicted and measured 
decay width, which lead to lively debates on its significance 
vis-a-vis the {\em experimental} error. No concern was expressed about 
the fact that the $Z^0$ width was calculated on the quark-gluon 
level, yet measured for hadrons. Likewise the strong coupling 
$\alpha_S(M_Z)$ is routinely extracted from the 
perturbatively computed hadronic $Z^0$ width with a stated 
theoretical uncertainty of 0.003 with translates into a 
theoretical error in $\Gamma _{had}(Z^0)$ of about 0.1\%.

There  are, however, several different versions and implementations of
the concept of duality. The problem with invoking duality implicitly is
that it is  very often unclear which version is used. In $B$ physics --
in  particular when determining $|V(cb)|$ and $|V(ub)|$ -- the 
measurements have become so precise that theory can no longer 
hide behind experimental errors. To estimate theoretical 
uncertainties in a meaningful way one has to give clear meaning 
to the concept of duality; only then can one analyze its 
limitations.

In response to the demands of $B$ physics a considerable literature 
has been created on duality over the last few years, which we want to
summarize. We will emphasize and illustrate the  underlying principles;
technical details can be found in the references we list.

Duality for processes involving time-like momenta was first addressed
theoretically in the late  '70's in references \cite{PQW76} and
\cite{GRECO}. We sketch  here the argument of Poggio, Quinn and Weinberg
since it contains  several of the relevant elements in a nutshell. The
cross section  for $e^+e^- \to hadrons$ can be expressed 
through an operator product expansion (OPE) of two hadronic 
currents.  One might be tempted to think that by invoking 
QCD's asymptotic freedom one can compute 
$\sigma (e^+e^- \to {\rm hadrons})$ for large c.m. energies 
$s \gg \Lambda _{QCD}$ in terms of quarks (and gluons) since 
it is shaped by short distance dynamics. However production 
thresholds like for charm induce singularities that vitiate 
such a straightforward computation. This complication can be 
handled in the following way. Consider 
the correlator of two electromagnetic currents: 
\begin{equation} \label{corr}
T_{\mu \nu}(q^2) = \int d^4 x \, e^{iqx} \, \langle 0 |
         T \left( J_\mu (x) J_\nu (0) \right) | 0 \rangle
= ( q_\mu q_\nu - g_{\mu \nu} q^2 ) \Pi(q^2) 
\end{equation}
where $\Pi(q^2)$ can be written in terms of a spectral function
$\rho(s)$ using an unsubtracted dispersion relation:
\begin{equation}
\Pi(q^2)  = \int \frac{ds}{2 \pi} \frac{\rho(s) }{q^2 - s + i \epsilon}
\end{equation}
It is well known that, to leading order in $\alpha_{em}$, 
$\rho(s)$ is related to the total cross section
for $e^+ e^- \to $ hadrons
\begin{equation}
\sigma(s) = \frac{4 \pi \alpha_{em}}{s} \rho(s) 
\end{equation}
Relying on QCD's asymptotic freedom one computes the 
correlator (\ref{corr}) in terms of quarks and gluons for 
$s$ in the deep Euclidean domain $|s| \gg \Lambda_{QCD}$; 
$s$ is chosen Euclidean so that one avoids a proximity to 
singularities induced by hadronic thresholds like for 
charm production etc. From the spectral function $\rho$ 
calculated in the Euclidean regime one can infer the cross 
section for physical, namely Minkowskian values of $s$. 
However, one cannot obtain it as a point-for-point function 
of $s$, only averaged -- or `smeared' -- over an energy interval, 
which can be written symbolically as follows:  
\beq 
\rho (s_{Euclid}) \; \Rightarrow 
\int _{s_0}^{s_0+\Delta s} ds\sigma (e^+e^- \to hadrons) 
\eeq 
This feature is immediately obvious: for the smooth $s$ dependence 
$\rho$ has to be compared to the measured
cross section $e^+ e^- \to$ hadrons as a function of $s$, which has
pronounced structures, in particular close to thresholds for $c \bar{c}$-
and $b\bar{b}$-production.

This simple illustration already points to the salient elements 
and features of duality and its limitations: 
\begin{itemize}
\item 
An OPE description in terms of quark and gluon degrees of freedom
for the observable under study is required.
\item 
This OPE has to be constructed in the Euclidean domain. 
\item 
Its results are analytically continued to the Minkowskian domain with 
the help of a dispersion relation. 
\item 
This extrapolation implies some loss of information; i.e. in the 
notation given above 
\begin{equation}
\langle T^{hadronic}_{\mu \nu} \rangle _w \simeq
          \langle T^{partonic}_{\mu \nu} \rangle _w
\label{ANSATZ} 
\end{equation}
where $\langle ... \rangle _w$ denotes the smearing which is
an average using a smooth weigth function $w(s)$; it generalizes 
the simplistic use of a fixed energy interval:
\begin{equation}
\langle ... \rangle _w = \int ds \, ... \, w(s)
\end{equation}

\item 
Some contributions that are quite insignificant in the Euclidean 
regime and therefore cannot be captured through the OPE can become 
relevant after the analytical continuation to the Minkowskian domain, 
as explained later on. 
For that reason we have used the approximate rather than the equality 
sign in Eq.(\ref{ANSATZ}).

\item 
One can make few universal statements on the numerical validity of 
duality. How much and what kind of smearing is required depends on the 
specifics of the reaction under study. 
\end{itemize}
The last item needs expanding right away. The degree to which 
$\langle T^{partonic}_{\mu \nu} \rangle _w$ can be trusted as a
theoretical description of the observable 
$\langle T^{hadronic}_{\mu \nu} \rangle _w $ depends on the weight 
function, in particular its width. It can be 
broad compared to the structures that may appear
in the hadronic spectral function, or it could be quite narrow, 
as an extreme case even $w(s) \sim \delta(s-s_0)$. It has become 
popular to refer to the first and second scenarios as 
{\em global} and {\em local} duality, respectively. Other authors 
use different names, and one can argue that this nomenclature 
is actually misleading. Below we will describe these items in 
more detail without attempting to impose ex cathedra a
uniform nomenclature ourselves.

Irrespective of names, a fundamental distinction concerning  
duality is often drawn between semileptonic and nonleptonic widths. 
Since the former neccessarily involves smearing 
with a smooth weight function due to the 
integration over neutrino momenta, it is often argued that predictions 
for the former are fundamentally more trustworthy than for the latter. 
However, as we shall see, such a categorical distinction is 
overstated and artificial; also it is  
not needed for the discussion in the following chapters. Of much 
more relevance is the differentiation between distributions and 
fully integrated rates.

No real progress beyond the more qualitative arguments of 
Refs. \cite{PQW76} and \cite{GRECO} occurred for many years. For as long 
as one has very limited control over nonperturbative effects, 
there is little meaningful that can be said about duality violations. 
Yet this has changed for heavy flavour physics with the development 
of heavy quark expansions, since within this OPE framework we can assess
nonperturbative effects as well as duality violations. 

The remainder of this note will be organized as follows:
 In the next section we shall give a more precise definition of what is
meant with ``Parton Hadron Duality'', which then allows us a discussion of
possible violations of duality in section~\ref{sec:violation}.
Based on this we give hints on how to check the concept of duality
in section~\ref{sec:validty}, before presenting conclusions.

%%%%%%%%%%%%%%%%%
\section{What is Parton--Hadron Duality?}
\label{DEF}
%%%%%%%%%%%%
%
%
In order to discuss possible violations of duality we have to give
first a more precise definition of this notion, which requires 
the introduction of some theoretical tools. We follow
closely the arguments given in the extensive reviews 
of Ref. \cite{shifman} and \cite{BU2001}\footnote{It can be
noted that even the authors of Ref.\cite{shifman} and \cite{BU2001}
-- although very close in the substance as well as the spirit
of their discussion -- do not use exactly the same terminology
concerning different aspects of duality.}. The central ingredient
into this definition is the method of the Wilsonian 
{\em Operator Product Expansion} (OPE) 
frequently used in field theory to
perform a separation of scales. In practical terms  this means that we
can write 
\begin{equation} \label{OPE}
\int d^4 x \, e^{iqx} \,
\langle A | T \left( J^\mu (x) J^\nu (0) \right) | A \rangle  
\simeq \sum_n \left(\frac{1}{Q^2} \right)^n   
c_n^{\mu \nu}(Q^2;\lambda)
  \langle A |{\cal O}_n | A \rangle _{\lambda} 
\end{equation}
for $Q^2 = -q^2 \to \infty$. The following notation has been used: 
$|A\rangle$ denotes a state that could be the  vacuum -- as for $e^+e^-
\to hadrons$ considered above -- or a 
$B$ meson when describing semileptonic beauty decays. $J^\mu$ 
denote electromagnetic and weak current operators for the former and 
the latter processes, respectively;  for other 
decays like nonleptonic or radiative ones one employs different 
$\Delta B =1$ operators; the ${\cal O}_n$ are local operators of 
increasing dimension. The operator of lowest dimension yields 
the leading contribution. In $e^+e^-$ annihilation 
it is the unit operator ${\cal O}_0 = 1$, for $B$ decays
${\cal O}_0=\bar bb$. They 
produce (among other things) the naive partonic results. Yet the 
OPE allows us to
systematically improve the naive partonic result. 
The coefficients $c_n^{\mu \nu}$ contain the 
contributions from short distance dynamics calculated
perturbatively based on QCD's asymptotic freedom. 
Following Wilson's prescription a mass scale $\lambda$ has been 
introduced to separate long and short distance dynamics; both the 
coefficients and the matrix elements depend on it, their product 
of course not.

The perturbative expansion takes the form
\begin{equation} \label{pert}
c_n^{\mu \nu}  = \sum_i \left(\frac{\alpha_S (Q^2)}{\pi} \right)^i
                 a_{n,i}^{\mu \nu}
\end{equation}
and is performed in terms of quarks and gluons.
The expectation values for the local operators provide the gateways 
through which nonperturbative dynamics enters.

The crucial point is that the OPE result is obtained in the 
Euclidean domain and has to be continued analytically into the 
Minkowskian regime relating the OPE result to observable hadronic 
quantities. As long as QCD is the theory of the strong interactions, 
it does not exhibit unphysical singularities in the complex $Q^2$ plane, 
and the analytical continuation will not induce additional contributions. 
To conclude: {\em duality between 
$\langle T^{hadronic}_{\mu \nu} \rangle _w$ and 
$\langle T^{partonic}_{\mu \nu} \rangle _w$ arises due to the existence 
of an OPE that is continued analytically}. It is just a restatement 
of QCD's basic tenet as the theory of the strong interactions that 
hadronic observables can be expressed in terms of quark-gluon 
degrees of freedom {\em provided} all possible sources of corrections 
to the simple parton picture are properly accounted for. 
It is thus misleading to refer to duality as an additional assumption.

Up to this point our discussion was quite generic. To specify it for 
semileptonic $B$ decays one chooses the current $J_{\mu}$ to be the 
weak charged current driven by $b\to c$ or $b \to u$. The expansion 
parameter for inclusive semileptonic decays is given by the 
energy release $\sim 1/(m_b - m_c) \; [1/m_b]$ for 
$b\to c \; [b\to u]$. For the exclusive mode $B \to l \nu D^*$ it is 
$1/m_b$ and $1/m_c$ with the latter yielding the numerically 
leading contributions.

%%%%%%%%%%%
\section{Duality Violations and Analytic Continuation} 
\label{sec:violation}
%%%%%%%%%
%
%
One of the main applications of the heavy quark expansion
is the reliable extraction 
of $|V(cb)|$ and $|V(ub)|$. One wants to be able to arrive at a 
meaningful estimate of the theoretical uncertainty in the values 
obtained. There are three obvious sources of theoretical 
errors: 
\begin{enumerate}
\item 
unknown terms of higher order in $\alpha_S$; 
\item 
unknown terms of higher order in $1/m_Q$;
\item 
uncertainties in the input parameters $\alpha_S$, $m_Q$ 
and the expectation values.  
\end{enumerate}
Duality violations constitute uncertainties {\em over and above} 
these; i.e. they represent contributions not accounted for due to 
\begin{itemize}
\item 
truncating these expansions at finite order and 
\item 
limitations in the algorithm employed.
\end{itemize}
These two effects are not unrelated. The first one means that the 
OPE in practice is insensitive to contributions of the type 
$e^{-m_Q /\mu}$ with $\mu$ denoting some hadronic scale; the second 
one reflects the fact that under an analytic continuation the 
term $e^{-m_Q /\mu}$, which is quite irrelevant for $Q = b$ -- 
though not neccessarily for $Q=c$ ! -- turns into an oscillating 
rather than suppressed term sin$(m_Q /\mu)$.

Of course we do not have (yet) a full theory for duality and its 
violations. Yet we know that without an OPE the question of 
duality is ill-posed. 
Furthermore in the last few years we have moved beyond the 
stage, where we could merely point to folklore. This progress has 
come about for the following reasons:  
\begin{itemize}
\item
We have refined our understanding of the physical origins of 
duality violations as due to 
\begin{itemize}
\item 
hadronic thresholds; 
\item 
so-called `distant cuts'; 
\item 
the suspect validity of $1/m_c$ expansions.
\end{itemize}
\item 
We understand the mathematical portals through which duality 
violations can enter, namely that the innocuous Euclidean 
quantity $e^{-m_Q /\mu}$ transmogrifies itself into the much 
more virulent Minkowskian quantity $\sin (m_Q /\mu)$ under 
analytical continuation. 
\item The quantity $e^{-m_Q/\mu}$ is actually innocuous for beauty,
   yet not necessarily for charm quarks. The `Euclidean' quantity
   $F_{D^*}(0)$ -- the formfactor for $B\to l \nu D^*$ at zero recoil --,
   which is given by an expansion in $1/m_c$, could be vulnerable to such
   a duality violation. However, the heavy mass expansion for exclusive
   quantities such as $F_{D^*}(0)$ is not directly given by an OPE, thus
   this argument may not apply in this case.
\item 
We have come up with an increasing array of field-theoretical 
toy models, chief among them the `t Hooft model, which is 
QCD in 1+1 dimensions in the limit of $N_C \to \infty$. It is 
solvable and thus allows an unequivocal comparison of the OPE 
result with the exact solution. 
\item 
For the analysis of $b\to c$ transitions we also have the small-velocity
expansion as a powerful tool.

\end{itemize}
We will not go into any details here, since they 
can be found in the literature \cite{BU2001}. The models do exhibit 
duality violations, but only highly suppressed ones conforming 
to general expectations. There had been claims of sizeable 
duality violations in the previous literature; those have been analyzed
carefully in Ref.\cite{BU2001}, where their flaws are pointed out
explicitly.

Based on general expectations as well as on analyzing the models 
one finds that indeed duality violations are described by highly 
power suppressed `oscillating' terms of the form 
\beq \label{DV}
T(m_Q) \sim \left( \frac{1}{m_Q}\right)^k {\rm sin}(m_Q\lambda) 
\eeq
for some integer power $k$. More generally one can state: 
\begin{itemize}
\item 
The primary criterion for addressing duality violation is the 
existence of an OPE for the particular observable. 
\item 
Duality will not be exact at finite masses. It represents an
approximation the accuracy of which will increase with the 
energy scales in a way that depends on the process in question. 
\item 
Limitations to duality can enter only in the form of an
oscillating  function of energy or $m_Q$ (or have to be exponentially
suppressed).  Duality violations {\em cannot be blamed for a 
systematic excess or deficit in the decay rates.} For example, no 
duality violation can convert $m_Q$ into $M_{H_Q}$ in the 
full width parametrically, only for discrete values of $m_Q$. 
\item 
The OPE equally applies to semileptonic as well as nonleptonic 
decay rates. Likewise both widths are subject to duality violations. 
The difference here is quantitative rather than qualitative; at finite 
heavy quark masses corrections are generally expected to be
larger in the nonleptonic 
widths. In particular, duality violations there can be boosted by the 
accidental nearby presence of a narrow hadronic resonance. Similar 
effects could arise in semileptonic rates, but are expected to be 
highly suppressed there. 
\item 
It is not necessary to have a proliferation of decay channels 
to reach the onset of duality, either approximate or asymptotic.  
Instructive examples are provided by the so-called small-velocity 
kinematics in semileptonic decays and by nonleptonic rates in the 
't Hooft model.
%\item Another subtlety could arise: one could conceivably have two
%   scales for the effective onset of duality -- a lower one when mainly
%   quark degrees of freedom are involved, and a higher one when gluonic
%   degrees of freedom become essential.
\end{itemize}

Putting everything together it has been estimated with considerable 
confidence -- at least by the authors of Ref.\cite{BU2001} -- 
that {\em duality violations in the integrated semileptonic width of 
$B$ mesons cannot exceed the fraction of a percent level. }
%\beq 
%\delta _{DV}\Gamma _{SL}(B) \leq {\rm few} \times 10^{-3} 
%\Gamma _{SL}(B)
%\eeq
As such we do not envision it to ever become the limiting 
factor in extracting $|V(cb)|$ and $|V(ub)|$ since the uncertainties 
in the expression for the semileptonic width due to fixed 
higher order contributions will remain larger than this level. 
The oscillatory nature of duality violating contributions is a main 
ingredient in this conclusion. It also shows that duality violations 
could become quite sizeable if an only partially integrated width 
-- let alone a distribution -- is considered. Generally, for distributions
the expansion parameter is not the heavy mass, rather it is a quantity 
such as $1/[m_Q (1-x)]$ where $x$ is e.g. the rescaled charged lepton
energy of a semileptonic decay. From equation (\ref{DV}) one would expect
that contributions the form $\sin(m_Q [1-x])/[m_Q (1-x)]^k$ appear in
differential distributions.

%%%%%%%%%%%%
\section{How can we check the validity of Parton--Hadron Duality?}
\label{sec:validty}
%%%%%%%%%%%
%
%

%We can fully understand that skeptics will adhere to Lenin's 
%dictum that trust is good, yet control is better.
If in the 
future we were to find a discrepancy between the measured and 
predicted values for, say, a CP asymmetry in $B$ decays, we had to check 
very diligently all ingredients upon which the prediction was
 based, in particular the values for $V(cb)$ and $V(ub)$, before 
we could make a credible claim to have uncovered New Physics. This 
means one needs a measure for potential duality violations that is 
not based purely on theoretical arguments.

Most theoretical uncertainties are systematical rather than statistical.  
As it is the case for experimental systematics the most convincing 
way to establish control over them is to determine the same quantity 
in independent ways and analyze their consistency.
The heavy quark expansions lend themselves 
naturally to such an approach since it is their hallmark that 
they allow the description of numerous decay rates in terms 
of a handful of basic parameters, namely quark masses and hadronic 
expectation values. Again the situation is very similar as for the 
perturbative series: once the coupling constant is determined (e.g.
$\alpha_S (M_Z)$) from a measurement, 
one may use this as an input to all other perturbative calculations,
thereby predicting other measurements.
If a prediction obtained in this way fails, one would
conclude that higher order effects have to be unusually large 
or that there is another deeper reason why a perturbative treatment 
does not apply.

Such independent determinations of the same quantity of course 
probe the overall theoretical control that we have established. By 
themselves they do not tell us whether a failure found is due to 
unusally large higher order contributions or to a breakdown in duality.

The fact that both the inclusive and exclusive methods for extracting 
$|V(cb)|$ yield consistent values -- and that the theoretical 
corrections one had to apply are both nontrivial and
essential for the agreement -- is such a
test.  We want to point to two other such tests that have become
available, namely concerning the $b$ quark mass and its kinetic energy 
expectation value.

\subsection{$b$ quark mass}

The $b$ quark mass has been extracted from beauty production at threshold
in $e^+e^-$ annihilation by several authors \cite{KUEHN}. Their findings 
can be stated in terms of two definitions of quark masses:
\begin{itemize}
\item[(i)] The  `kinetic mass' is defined by 
\beq 
\frac{dm_Q^{kin}(\mu)}{d\mu} = 
-\frac{16}{9}\frac{\alpha _S(\mu)}{\pi} - 
\frac{4}{3} \frac{\alpha _S(\mu)}{\pi} \frac{\mu}{m_Q} 
+ {\cal O}\left( \alpha _S^2,\alpha _S\cdot \frac{\mu^2}{m_Q^2}\right) 
\; , 
\eeq
normalized at 1 GeV; it is well-defined in full QCD and does not 
suffer from a renormalon ambiguity; equivalently one can use  
\beq \label{lambdadef}
\bar \Lambda (\mu) \equiv {\rm lim}_{m_Q \to \infty} 
\left[ 
M(H_Q) - m_Q^{kin}(\mu) 
\right] \; .
\eeq
where $M (H_Q)$ is the mass of the $0^-$ ground state.
\item[(ii)]
The pole or HQET mass which is a very popular choice, although it is 
not well-defined in full QCD since it suffers from the renormalon ambiguity. 
If appropriate care and caution are applied one can still use it in 
calculation; as a rule of thumb one has for its relatioship to the 
kinetic mass: 
\beq 
\bar \Lambda _{\rm HQET} = \bar \Lambda (1\; {\rm GeV}) - 
0.255 \; \GeV
\eeq
where the parameter $\bar\Lambda$ is defined in the same way as in
(\ref{lambdadef}):
\beq
\bar \Lambda ^{HQET} \equiv {\rm lim}_{m_Q \to \infty}
[M(H_Q) - m_Q^{pole}]
\eeq
\end{itemize}
The results  
are completely consistent within the stated uncertainties of 
1-2 \% and can be summarized as follows: 
\beq 
m_b^{kin}(1\; {\rm GeV})|_{e^+e^-\to \bar bb} = 4.57 \pm 0.06 \; {\GeV}
\;  
\leftrightarrow  \; \bar \Lambda (1\GeV) = 0.71 \pm 0.06 \; \GeV 
\label{MKINAV}
\eeq 
or 
\beq 
m_b^{pole} = 4.82 \pm 0.06 \; \GeV \; 
\leftrightarrow  \; \bar \Lambda ^{HQET} = 0.45 \pm 0.06 \; \GeV
\eeq

The techniques employed in the analysis differ 
somewhat from author to author; the full agreement in their findings 
is thus quite re-assuring. One should keep in mind, though, 
that these determinations share their experimental input to a large 
degree. The value stated in Eq.(\ref{MKINAV}) could thus be subject to 
some systematic shift from the true value. Arguments based on the 
small-velocity sum rules indeed suggest that 
$m_b^{kin}(1\; {\rm GeV})$ could lie a bit above 4.6 GeV.

\begin{figure}[hbt]
\begin{center}
\epsfig{scale=0.5,file=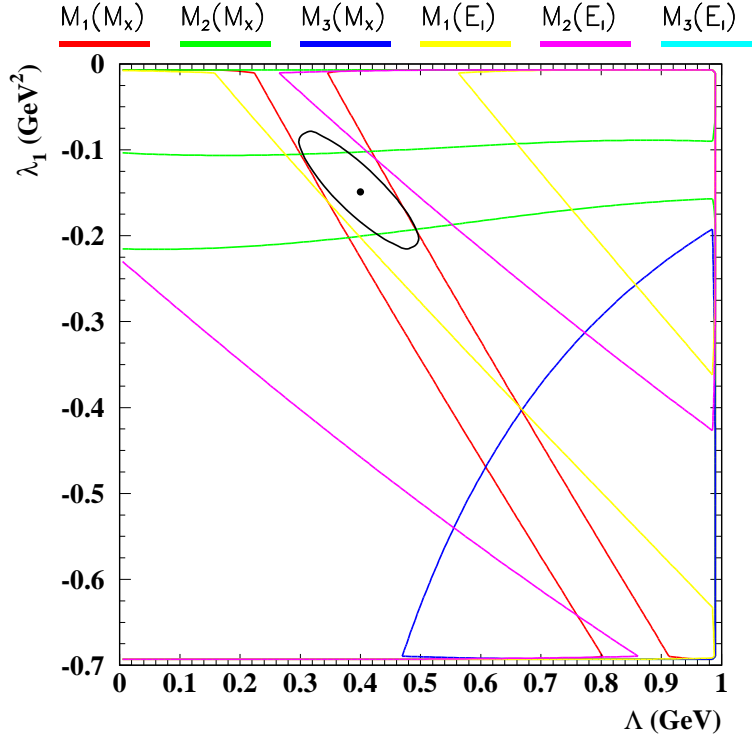}
\end{center}
\caption{Contours in the $\bar\Lambda$-$\lambda_1$ plane,
         from the DELPHI measurement of the first and second
moments \cite{DELPHI}. The band defined by the almost flat
curves in the upper part are from the second moment of the hadronic
invariant mass, the steepest straight line is from the first moment of the
hadronic invariant mass. The remaining curves are from the moments of the
lepton energy, the steeper being from the first moment. The curve in the
lower right is from the third moment of the hadronic in variant mass.}
\label{fig1}
\end{figure}

\begin{figure}[hbt]
\begin{center}
\epsfig{scale=0.6,file=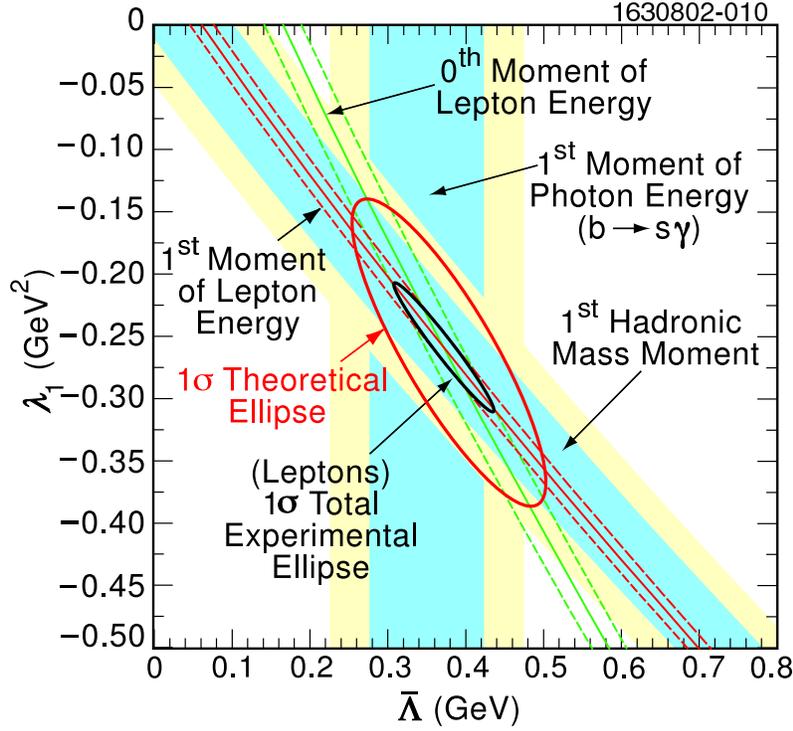}
\end{center}
\caption{Contours in the $\bar\Lambda$-$\lambda_1$ plane,
         from the CLEO  measurements of the moments \cite{Briere:2002hw}.}
\label{fig2}
\end{figure}

The $b$ quark mass also affects the shape 
of lepton energy and hadronic mass spectra in semileptonic (and 
photon spectra in radiative) $B$ decays. Its value can therefore
be obtained from the measured lepton energy
   and hadronic mass moments, which encode the shape of these spectra.
The DELPHI and CLEO
collaborations have presented data as shown in the Figures~\ref{fig1}
and \ref{fig2}. 
Again it is pleasing to see that the different moments indeed 
yield completely consistent values although this is not truly surprising
since they are highly correlated. DELPHI finds 
\bea 
m_b^{kin}(1\; {\rm GeV})|_{mom} &=& 4.59 \pm 0.08 \pm 0.01
\; \GeV  \\ \nonumber 
&\leftrightarrow&  \; \bar \Lambda (1\GeV) = 0.69 \pm 0.08 \pm 0.01
\; \GeV  \\
m_b^{pole}|_{mom} &=& 4.88 \pm 0.10 \pm 0.02 \; 
\GeV \\ \nonumber
&\leftrightarrow&  \; \bar \Lambda ^{HQET} = 0.40 \pm 0.10 \pm 0.02 
\; \GeV
\eea
It is again reassuring that their
fit results are consistent with Eq.(\ref{MKINAV})
CLEO measures truncated lepton
   energy moments and states their findings in terms of HQET
   parameters
\bea 
m_b^{pole}|_{mom} &=& 4.88 \pm 0.03 \pm 0.06 \pm 0.12 \; 
\GeV \\ \nonumber 
&\leftrightarrow&  \; \bar \Lambda ^{HQET} = 0.39 \pm 0.03 \pm 0.06 \pm 0.12
\; \GeV
\eea

The main news here is how well the values for 
$m_b$ extracted from two sources agree that are quite different in their 
experimental as well as theoretical systematics, namely weak
$B$ decays and the electromagnetic production of beauty at threshold.

\subsection{Average kinetic energy}

A similarly pleasing picture emerges from determining the 
kinetic expectation value. Based on QCD sum rules and 
SV sum rules one had inferred for the infrared
stable quantity 
$\mu _{\pi}^2(1 \GeV)$: 
\beq 
\mu _{\pi}^2(1 \GeV) \simeq 0.45 \pm 0.1 \; (\GeV)^2 \; . 
\eeq 
Using the rule of thumb for the relation to the HQET parameter 
$\lambda_1$ \cite{Uraltsev:2002ta}

\beq 
- \lambda _1 \simeq \mu _{\pi}^2(1 \GeV) - 0.18 (\GeV)^2
\eeq 
this estimate translates into 
\beq 
- \lambda _1 \simeq 0.27 \pm 0.1 (\GeV)^2
\eeq
The aforementioned DELPHI and CLEO analyses also yield values 
for this quantity, namely 
\bea 
\mu _{\pi}^2(1 \GeV) &=& 0.31 \pm 0.07 \pm 0.02 \; (\GeV)^2 
\; \; {\rm DELPHI}\\ 
- \lambda _1 &=& 0.15 \pm 0.07 \pm 0.03  \; (\GeV)^2 
\; \; {\rm DELPHI} \\
- \lambda _1 &=& 0.25 \pm 0.02 \pm 0.05 \pm 0.14  \; (\GeV)^2 
\; \; {\rm CLEO}
\eea

The fact that the parameters extracted in different way and form different
observables yield consistent values for the quark mass and the kinetic energy
parameter indicates that no anomalously large higher order
corrections or unexpectedly
sizeable duality violating contributions are present.

\section{Conclusions}

From all what we know currently from purely theretical
considerations duality violations should be safely below one percent
in the semileptonic branching ratio. This is likely to remain in the noise
   level of theoretical uncertainties due to terms of order $1/m_b^3$
   and higher and of higher order perturbative contributions.
Hence we do not see any need to
assign some additional uncertainty to the extraction of $V_{cb}$ from
a possible duality violation in inclusive decays.
   This should not be seen as an
   ex cathedra statement. When more and more types of moments will be
   measured with more and more accuracy -- even separately in
   the decays of $B_d$, $B^-$ and $B_s$ mesons --, additional
   constraints will be placed on the same set of heavy quark
   parameters. This will provide highly nontrivial tests of our
   theoretical control.

\section*{Acknowledgements}
We gratefully acknowledge illuminating discussions with N. 
Uraltsev. This work has been supported by the National Science Foundation 
under grant number PHY00-87419, bt the DFG Research Group
``Quantenfeldtheorie, Computeralgebra und Montre Carlo Simulationen'' and
by the German Minister for Education and Research BMBF und grant number
05HT1VKB1 and the DFG Merkator Program.


\begin{thebibliography}{99}
\bibitem{PQW76}
E.~C.~Poggio, H.~R.~Quinn and S.~Weinberg,
%``Smearing The Quark Model,''
Phys.\ Rev.\  {\bf D 13}, 1958 (1976).
%%CITATION = PHRVA,D13,1958;%%
\bibitem{GRECO}
M. Greco, G. Penso and V. Srivastava, Phys.\ Rev.\ {\bf  D 12} (1980) 2520.
\bibitem{shifman}
M.~A.~Shifman, ``Quark-hadron duality,''
published in the Boris Ioffe Festschrift
'At the Frontier of Particle Physics /
Handbook of QCD', ed. M. Shifman (World Scientific, Singapore, 2001), 
arXiv:hep-ph/0009131.
%%CITATION = HEP-PH 0009131;%%
\bibitem{BU2001}
I.~I.~Bigi and N.~Uraltsev,
%``A vademecum on quark hadron duality,''
Int.\ J.\ Mod.\ Phys.\  {\bf A 16}, 5201 (2001)
[arXiv:hep-ph/0106346].
%%CITATION = HEP-PH 0106346;%%
\bibitem{shifmanAsy}
M. Shifman, in {\it Continuous Advances in QCD}, Ed. A. Smilga
(World Scienific, Singapore 1994), p.249 [hep-ph/9405246],
and in {\it Particles, Strings and Cosmology} Eds. J. Bagger et al.,
(World Scientific, Singapore 1996) p.69 [hep-ph/9505289].
\bibitem{Isgur}
N.~Isgur,
%``Duality-violating 1/m(Q) effects in heavy quark decay,''
Phys.\ Lett.\ {\bf B 448}, 111 (1999)
[arXiv:hep-ph/9811377].
%%CITATION = HEP-PH 9811377;%%
\bibitem{LY}
A.~Le Yaouanc, D.~Melikhov, V.~Morenas, L.~Oliver, O.~Pene and J.~C.~Raynal,
Phys.\ Lett.\  {\bf B 517}, 135 (2001)
[arXiv:hep-ph/0103339].
%%CITATION = HEP-PH 0103339;%%
\bibitem{Grinstein}
B.~Grinstein and R.~F.~Lebed,
%``Explicit quark-hadron duality in heavy-light meson weak decays in the  't 
% Hooft model,''
Phys.\ Rev.\  {\bf D 57}, 1366 (1998)
[arXiv:hep-ph/9708396].
%%CITATION = HEP-PH 9708396;%%
\bibitem{KUEHN}
K. Melnikov and A. Yelkhovsky, Phys.\ Rev.\ {\bf D 59} (1999) 114009; \\
    A. Hoang, Phys.\ Rev.\ {\bf D 61} (2000) 034005; \\
    M. Beneke and A. Signer,
    Phys.\ Lett.\ {\bf  B 471} (1999) 233; \\
    .~H.~Kuhn and M.~Steinhauser,
Nucl.\ Phys.\  {\bf B 619}, 588 (2001)
[Erratum-ibid.\  {\bf B 640}, 415 (2002)]
[arXiv:hep-ph/0109084].
%%CITATION = HEP-PH 0109084;%%
\bibitem{Moments}
P.~Gambino, these processdings
%
\bibitem{DELPHI} 
M.~Battaglia {\it et al.},
%``Heavy quark parameters and $|$V(cb)$|$ from spectral moments in
% semileptonic B decays,''
CERN-TH/2002-290, arXiv:hep-ph/0210319.
%%CITATION = HEP-PH 0210319;%%...
\bibitem{Briere:2002hw}
R.~A.~Briere {\it et al.}  [CLEO Collaboration],
Contributed to 31st International Conference on High Energy
Physics (ICHEP 2002), Amsterdam, The Netherlands, 24-31 Jul 2002,
%``Measurement of the lepton energy in the decay anti-B $\to$ X l anti-nu  and 
determination of the heavy quark expansion parameters,''
arXiv:hep-ex/0209024.
%%CITATION = HEP-EX 0209024;%%
%
\bibitem{Uraltsev:2002ta}
N.~Uraltsev, Talk given at 31st International Conference on
High Energy Physics (ICHEP 2002), Amsterdam, The
Netherlands, 24-31 Jul 2002,  
%``Strong interaction effects in semileptonic B decays,''
arXiv:hep-ph/0210044.
%%CITATION = HEP-PH 0210044;%%
\end{thebibliography}
\end{document}